\documentclass[twocolumn,aps,prr,superscriptaddress,10pt,biblography]{revtex4-2} 

\usepackage{dcolumn}
\usepackage{amsmath}
\usepackage{multirow}
\usepackage{mathtools}
\usepackage{bbold}
\usepackage{hyperref}
\hypersetup{colorlinks=true,allcolors=blue}
\usepackage{tikz}
\begin{document}

\title{Universal Topological Power Transfer with Arbitrarily Large Chern Number \\ in Driven Quantum Spin Chains}

\author{Anshuman Tripathi}
\affiliation{I.~Institut f{\"u}r Theoretische Physik, Universit{\"a}t Hamburg, Hamburg, Germany}
\affiliation{The Hamburg Centre for Ultrafast Imaging, Hamburg, Germany}

\author{Mircea Trif}
\affiliation{International Research Centre MagTop, Institute of Physics, Polish Academy of Sciences, Warsaw, Poland}

\author{Thore Posske}
\affiliation{I.~Institut f{\"u}r Theoretische Physik, Universit{\"a}t Hamburg, Hamburg, Germany}
\affiliation{The Hamburg Centre for Ultrafast Imaging, Hamburg, Germany}

\begin{abstract}
\noindent Topological frequency converters exploit a quantized transfer of power between two driving fields in a quantum system, a phenomenon topologically protected by the Chern number of the associated fiber bundle. While realizations with few-spin systems have theoretically demonstrated this effect, the conversion factors have typically been restricted to small integer values. Here, we investigate an interacting $XXZ$ Heisenberg spin-$1/2$ chain driven adiabatically by two magnetic drives with incommensurate frequencies. The Chern number, determined by the degeneracy points enclosed by the adiabatic trajectory, increases systematically with the chain length and can be tuned through the exchange anisotropy, providing direct control of the topological pumping strength. We reveal a universal dependency of the anisotropy and magnetic field strength for odd and even chain lengths of the quantum critical points. This provides a mechanism for a topological frequency converter with an arbitrarily large, quantized conversion ratio in interacting quantum spin chains. The mechanism remains the same for arbitrary coupling regions of the drives. 
\end{abstract}

\maketitle

\section{Introduction}
\noindent The concept of topological pumping originally demonstrates that the adiabatic cyclic evolution of a one-dimensional quantum system can result in quantized transport of charge determined by the topology of the Hamiltonian \cite{thouless_quantization_1983, shindou_quantum_2005, citro_thouless_2023, fu_time_2006}. Such systems, when driven by multiple incommensurate frequencies that introduce synthetic dimensions, lead to quantized power transfer, known as topological frequency conversion between the drives \cite{martin_topological_2017,crowley_topological_2019,korber_interacting_2020,brouwer_scattering_1998, crowley_half-integer_2020, nathan_topological_2019, ozawa_topological_2019, casati_anderson_1989}. This is understood through Floquet lattice theory, where the phases of the drives act as the synthetic dimension, and the topological properties of the resulting fiber bundle lead to the Chern number $C$ as the relevant integer topological invariant \cite{martin_topological_2017,lindner_floquet_2011, zhenghao_floquet_2011, Grifoni_driven_1998, Gomez_Floquet_2013, cayssol_floquet_2013, ozawa_topological_2019}. If a system is driven by two incommensurate frequencies, $\omega_1$ and $\omega_2$, the power transfer follows the relation $P \propto C \omega_1 \omega_2$ \cite{martin_topological_2017,crowley_topological_2019}. However, current suggestions for systems typically limit the Chern number to low integer values, which restricts the amplification factor of these converters \cite{korber_interacting_2020, crowley_half-integer_2020, crowley_topological_2019, martin_topological_2017}.    

In this work, we first establish a realization with an arbitrarily large Chern number by using a Heisenberg $XXZ$ spin-$1/2$ chain driven adiabatically at its boundaries. When driven by two incommensurate circularly polarized periodic magnetic drives, the dynamics correspond to a trajectory on a self-intersecting torus defined by the phases of the boundary drives. The topological nature of the pump is determined by the specific degeneracy points of the ground state that are enclosed by the torus. We uncover that the number of these enclosed singularities, and consequently the Chern number, increases monotonically with the length of the spin chain, with a linear increase for anisotropy values close to $\Delta/J =1$, allowing for the dynamical tuning of the quantized transfer. We analyze the Chern number in dependence on the anisotropy and the magnitude of the driving fields, determining the tuning parameters. The shape of the gap closing parameters, i.e., the quantum critical points, follows universal curves in parameter space, yet differentiates between odd and even chain lengths. This setup realizes a topological frequency converter where the conversion ratio is no longer limited but can be arbitrarily increased by extending the system size or adjusting interaction parameters. We analyze the relationship between the chain length, anisotropy, and the resulting Chern number, establishing a direct route to realizing high-gain topological frequency conversion in an interacting quantum system. Furthermore, we introduce an extension of this protocol where the driving fields are spatially distributed across both ends of the chain, yielding the same quantized pumping effect. This distributed configuration demonstrates that energy can be transported and converted along the entire chain by a topologically protected mechanism. 

\begin{figure*}[t]
        \centering
        \setlength{\unitlength}{0.1\textwidth}
        \begin{picture}(9.5,4)
            \put(0,0){\includegraphics[width=0.9\textwidth]{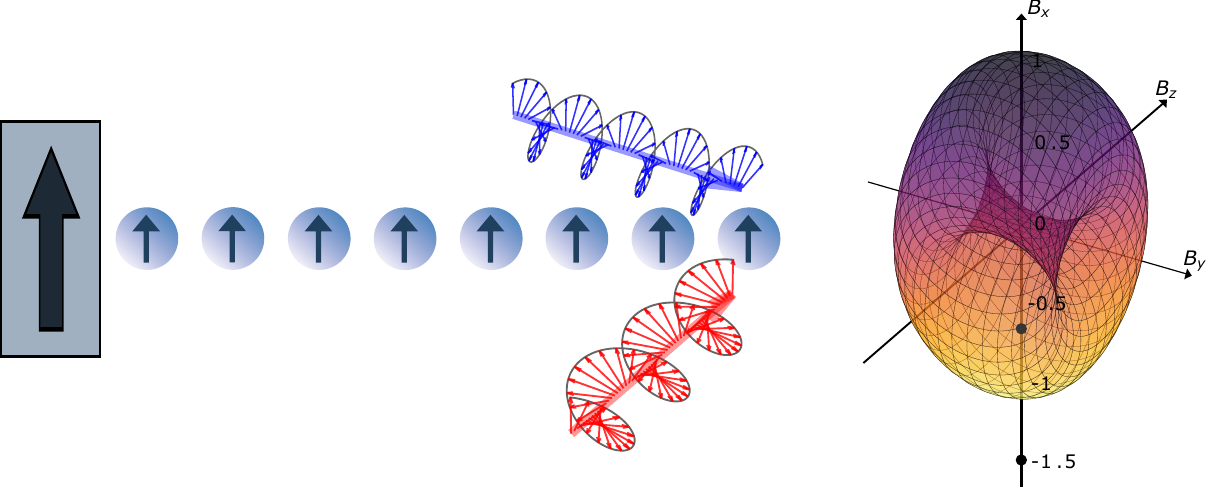}}
            \put(0.03,3.8){\textbf{(a)}}
            \put(6.5,3.8){\textbf{(b)}}
        \end{picture}        
    \caption{The frequency converter setup. (a) The $XXZ$ spin-$1/2$ chain, where the first spin is coupled to a static magnetic reservoir pointing in $x$-direction, and the last spin is coupled to the two incommensurate circularly polarized magnetic drives in the $xz$ (blue field) and $xy$ (red field) planes. (b) The self-intersecting torus traced by the drives for field strength $B = 1/2$. The black dots mark two degeneracy points, the one enclosed by the torus yields the Chern number $C= 1$ for a chain length $N = 8$ and anisotropy $\Delta = -0.78J$.}  
    \label{fig:Setup}
\end{figure*}

The outline of the manuscript is as follows. In Sec.~\ref {sec:Model}, we introduce the $XXZ$ model with boundary drives and the mechanism of power transfer between them. In Sec.~\ref{sec:Results}, we present the resulting topological phase diagram, highlighting the linear increase of the Chern number with system size and the universal collapse of the quantum critical points. We verify this power transfer for an example case by simulating the real-time evolution of the ground state across the quasi-adiabatic regime and comparing it with the diabatic regimes. We then generalize this setup, demonstrating that analogous topological transfer occurs under alternative driving and boundary field configurations. Finally, in Sec.~\ref{sec:Conclusion}, we conclude our findings, and highlight potential experimental realizations using magnetic adatom chains assembled via scanning tunneling microscopy \cite{khajetoorians_current-driven_2013, baumann_electron_2015, ternes_spin_2015, ChoiHeinrich_Atomic_2019, yang_probing_2021} and ultracold atoms in optical lattices with Feshbach resonance control of the Heisenberg anisotropy \cite{citro_thouless_2023, jepsen_spin_2020, trotzky_controlling_2010,chen_controlling_2011, weitenberg_single-spin_2011}.

\section{Model and Methods} \label{sec:Model} 
\noindent We investigate the frequency converter in an interacting XXZ Heisenberg spin-$1/2$ chain composed of $N$ quantum spins $\mathbf{S}_j$, where we fix one end of the chain with a static external magnetic field in the $x$-direction and adiabatically influence the other end with two mutually perpendicular rotating magnetic fields acting as phase drives, see Fig.~\ref{fig:Setup}. The effective Hamiltonian reads
\begin{align} \label{eq:model}
    H(\boldsymbol{\theta}_t) = & - \sum_{j=1}^{N-1}\left[J(S_j^{x} S_{j+1}^{x} +S_{j}^{y} S_{j+1}^{y}) + \Delta S_{j}^{z} S_{j+1}^{z}\right]  \nonumber \\
    &- J\left[\mathbf{B}_1\cdot\mathbf{S}_1 +  \mathbf{S}_{N}\cdot\mathbf{B}_{N}(\boldsymbol{\theta}_t)\right].
\end{align}
Here, we consider $\hbar = 1$. The external magnetic field at the first site $\mathbf{B}_1 = (1,0,0)$ is understood as the effect of a large field that fixes the quantum spin of one end of the chain, which we integrate out \cite{tripathi_generalized_2025}. The last site $\mathbf{B}_N(\boldsymbol{\theta}_t) = B(\cos{\theta_{1t}}, \sin{\theta_{1t}}, 0) + B(\cos{\theta_{2t}}, 0, \sin{\theta_{2t}})$ is affected by the two independent phase drives rotating in $xy$ plane and $xz$ plane respectively. The time dependence of the phase drives is $\boldsymbol{\theta}_t = (\theta_{1t} , \theta_{2t}) = (\omega_1t+\phi_1, \omega_2t +\phi_2)$ where $\omega$ and $\phi$ parametrize the frequency and offset, see Fig.~\ref{fig:Setup}. 

Under the influence of two incommensurate driving frequencies, $\omega_1$ and $\omega_2$, the  time-averaged energy pumping rate from one drive to the other becomes topologically quantized, see Appendix \ref{app:powerTransfer}, following the relation \cite{martin_topological_2017,crowley_topological_2019},
\begin{align}
    P = &\lim_{T \to \infty} \frac{1}{T} \int_{0}^{T} \left\langle \frac{\partial H}{\partial \theta_1} \right\rangle \omega_1 dt = (C/2\pi) \omega_1 \omega_2. 
\end{align}
 Here, $C$ represents the Chern number, which is the integral over the Berry curvature of the gapped ground state $|\psi_0(\mathbf{\theta})\rangle$ defined over the toroidal parameter space,
\begin{align} \label{Eq:ChernNumber}
    C = \frac{i}{2\pi} \iint_{0}^{2\pi}\left( \left\langle \frac{\partial \psi_0}{\partial \theta_1} \Bigg | \frac{\partial \psi_0}{\partial \theta_2} \right\rangle - \left\langle \frac{\partial \psi_0}{\partial \theta_2}\Bigg | \frac{\partial \psi_0}{\partial \theta_1} \right\rangle \right) {\rm d}\theta_1 {\rm d}\theta_2.
\end{align}
To evaluate this integral numerically, we discretize the parameter space $(\theta_1, \theta_2) \in [0,2\pi] \times [0,2\pi]$ into a regular grid and sum the local Berry fluxes for each elementary plaquette on this grid using the gauge-invariant projection method \cite{asboth_berry_2016}, where the ground states $|\psi_0(\mathbf{\theta})\rangle$ are obtained via the Lanczos algorithm \cite{brezinski_lanczos_1997}. Our calculations for chain lengths up to $N=11$ sites reveal that the Chern number equals the number of ground-state degeneracy points enclosed by the driving trajectory on the torus; i.e., each enclosed degeneracy point acts as a monopole, contributing a topological charge of exactly one.
    
\section{Results}\label{sec:Results}

\begin{figure*}[t]
        \centering
        \setlength{\unitlength}{0.1\textwidth}
        \begin{picture}(10,3.8)
            \put(0,0){\includegraphics[width=0.99\textwidth]{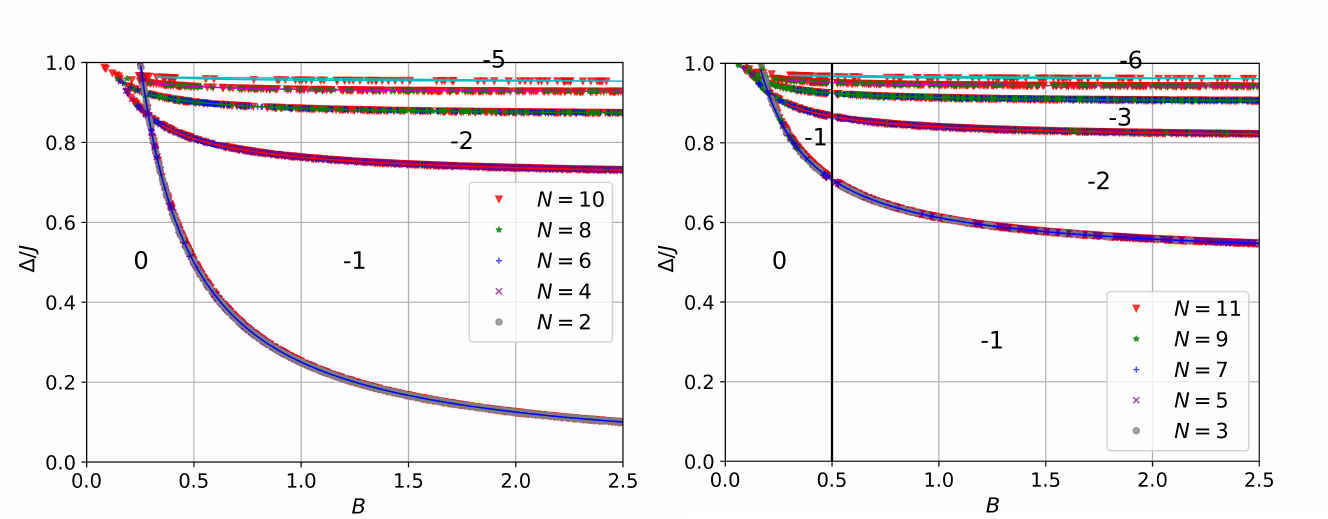}}
            \put(0.03,3.8){\textbf{(a)}}
            \put(5,3.8){\textbf{(b)}}
        \end{picture}        
    \caption{The topological sectors, separated by quantum critical points, i.e., parameters of gap closure $\delta E =0$. The universal collapse of trajectories as a function of anisotropy $\Delta$ and the magnetic field $B$ pointing in the $-x$ direction. The respective markers are the numerical value of the degeneracy calculated for (a) even and (b) odd chain lengths, and the colored curves are the hyperbolic fits $(\Delta/J - \cos(\pi/N))(B+b) = c$, see Appendix \ref{app:FitParameters} for values. The critical points accumulate in dependence on the chain length $N$ for both the odd and even cases. By choosing the magnitude and starting point of $B$, diverse topological pumps can be realized. }
    \label{fig:PhaseSpace}
\end{figure*}

\noindent Since the topological invariant is determined by the distribution of degeneracies, we map their landscape by analyzing the energy gap, $\delta E$, between the ground and first excited states. We define a four-dimensional parameter manifold spanned by the exchange anisotropy $\Delta$ and the boundary magnetic field drives $\mathbf{B}_N(\boldsymbol{\theta}_t)$ acting on the $N^{th}$ spin. To locate the singular points where $\delta E = 0$, we employ a numerical minimization protocol over the parameter bounds $ \Delta/J \in (0,1), B \in (0,2.5), \theta_1 \in (0,2\pi)$ and $ \theta_2 \in (0,2\pi)$, computing the gap at each iteration using the Lanczos algorithm. To ensure a comprehensive mapping and to find all gap-closings, we perform on the order of $10^4$ of gradient-based optimization runs, each initialized from a random starting point. This search reveals a collapse of the degeneracy loci: All spectral gap closings occur exclusively when the total boundary field $ \mathbf{B}_N $ aligns with the -$x$ direction, corresponding to $(\theta_1, \theta_2) = (\pi, \pi)$. This finding reduces the relevant phase space to the plane defined by the anisotropy $\Delta$ and the field magnitude $B$.

The topological sectors for up to the chain length of $N=11$ are presented in Fig.~\ref{fig:PhaseSpace}. The numerically obtained singularities $\delta E = 0$ collapse along universal curves independent of the length of the chain. We characterize these trajectories by fitting to $(\Delta/J - \cos(\pi/N))(B+b) = c$, where $b$ and $c$ are specific to each curve, see Appendix \ref{app:FitParameters} for concrete values. We observe that the curves saturate to $\cos{(\pi/N)}$, which is exactly the degeneracy point at $B = 1/2$ for chain length $N-1$ \cite{thore_winding_2019}. This is because the large magnetic field $B$ results in the fixation of the $N^{th}$ quantum spin to $(-1,0,0)$ in low energy, which effectively leads to a chain of $N-1$ quantum spins at exactly the degeneracy point \cite{tripathi_generalized_2025}. The agreement between the data points and this fitting function suggests a fundamental hyperbolic relationship between the exchange anisotropy and the critical magnetic field strength required to close the spectral gap. Even and odd chain lengths differ fundamentally in the topological structure. For systems with an even number of sites, the degeneracy curves appear successively from lower to larger anisotropies. For $N=2$, only the branch lowest in $\Delta$ exists. As the length increases to $N=4$, a second branch emerges at larger $\Delta$, and for $N=6$, a third branch is added to the stack, etc., each branch remaining to signify $\delta E=0$ for all longer chain lengths. This implies that the degeneracy curves are universal features that persist unchanged as the system grows. For systems with an odd number of sites, the degeneracy landscape exhibits a robust extra feature due to Kramers' degeneracy \cite{kramer_1930, klein_degeneracy_1952} at $B=1/2$ irrespective of the value of $\Delta$, illustrated by the vertical line in the phase diagram of Fig.~\ref{fig:PhaseSpace}(b). The pumping mechanism can therefore additionally be used as a detector for the parity of the chain length if $B > 1/2$. The remaining phenomenology is similar to the even $N$ case; additional hyperbolic branches accumulate at higher anisotropies as $N$ increases. However, these curves are distinct from those found in even-length chains, following a separate set of universal scaling parameters. Despite the parity-dependent differences in the specific locations of the degeneracy curves, the underlying mechanism for topological scaling remains identical. The hierarchical universal degeneracy curves are a central finding of this work.

\begin{figure*}[t]
        \centering
        \setlength{\unitlength}{0.1\textwidth}
        \begin{picture}(10,3.3)
            \put(0,0){\includegraphics[width=0.95\textwidth]{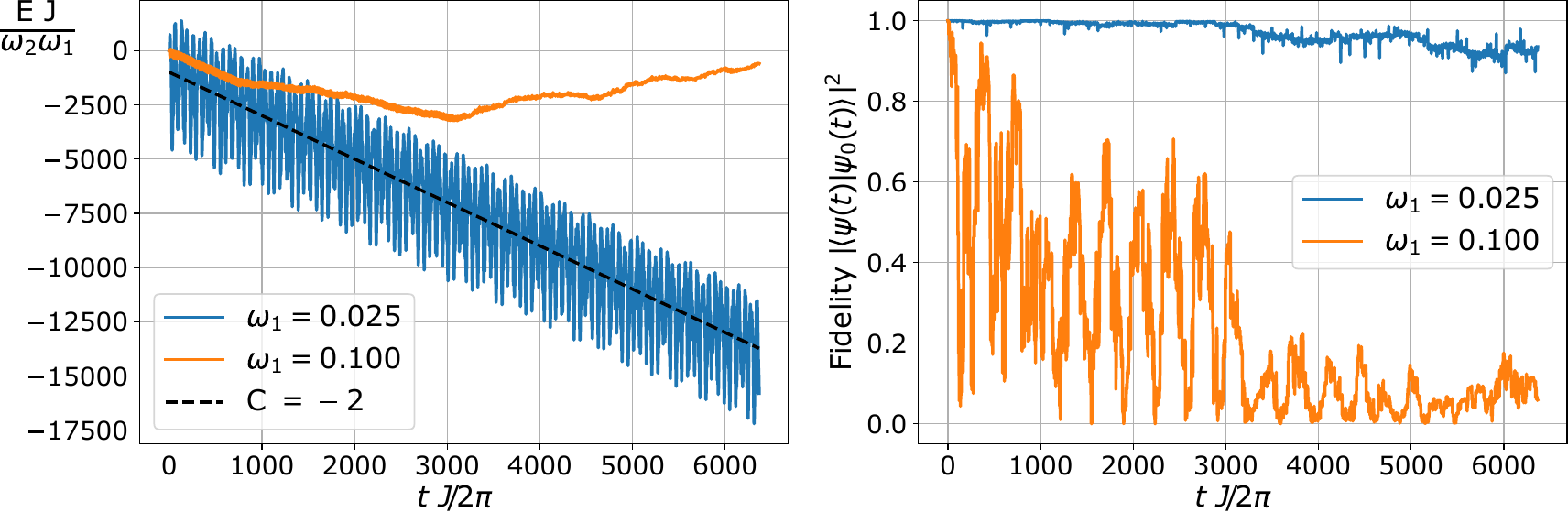}}
            \put(0.8,3.2){\textbf{(a)}}
            \put(5.5,3.2){\textbf{(b)}}
        \end{picture}        
    \caption{ Dynamics of energy transfer (a) and the corresponding ground-state fidelity (b). Time evolution is simulated for a spin chain of length $N=4$ with parameters $\Delta=J$ and $B=1$, driven by incommensurate frequencies related by the golden ratio, $\omega_2 = \omega_1(1+\sqrt{5})/2$. In the adiabatic regime ($\omega_1 = 0.025$), shown in the blue plot, the system remains near the instantaneous ground state, and the time-averaged slope of the energy transfer is quantized, corresponding to the Chern number $C=2$. In the diabatic regime ($\omega_1 = 0.1$), shown in the orange plot, transitions cause a significant deviation of fidelity from unity, leading to the breakdown of the linear topological pump. } 
    \label{fig:TimeEvolution}
\end{figure*}

This cumulative structure enables the linear scaling of the Chern number $C \propto N$. By operating at anisotropy close to $\Delta/J = 1$ and sufficiently strong $B$, the driving trajectory encloses all existing degeneracy branches, the number of which grows linearly with the chain length. Additionally, the separation of these curves enables dynamical tunability by varying $\Delta$ and $B$, which drives the system across consecutive degeneracy manifolds, triggering topological phase transitions. This allows switching between quantized pumping modes by adjusting the interaction strength and the strength of the driving field, especially accessible in the regime $\Delta/J < 1$. In the easy-axis regime with $\Delta/J>1$, the magnetic field amplitude $B$ of the degeneracies asymptotically saturates to the universal values $-b$, see Tab.~\ref{tab:Parameters}.

To demonstrate that the average power transfer between the driving tones is quantized and directly proportional to the Chern number, provided the evolution remains adiabatic, see Appendix \ref{app:powerTransfer}, we initialize the system in the ground state at $t=0$ and evaluate the time evolution $|\psi(t)\rangle$ by the Crank-Nicolson method \cite{Bauch_QuanutmPlasmaSimutationBook_2010}, ensuring that the norm of the wavefunction is precisely conserved over time. We monitor two primary observables during the evolution, the energy transfer and the fidelity $F(t)=| \langle \psi_0(t)| \psi(t) \rangle |^2$ that measures the overlap between the time-evolved state and the instantaneous ground state of the driven Hamiltonian. 

In Fig.~\ref{fig:TimeEvolution}, we show the pumped energy and fidelity for $N=4$ at $\Delta = J$ and $B=1$ for the quasiadiabatic case $\omega_1 = 0.025$ and for the diabatic case $\omega_1 = 0.1$ with the other incommensurate frequency set to $\omega_2 = \omega_1 (1+\sqrt{5})/2$, scaled by the golden ratio. For both cases, we set the offset phase to $\phi_1 = \phi_2 = 0$. For the quasiadiabatic $\omega_1 = 0.025$, the system exhibits a robust, linear transfer of energy over time. The evolution corresponds to the expected Chern number of $C= -2$. Throughout this process, the fidelity $F(t)$ remains above $0.85$, confirming that the driving frequencies are sufficiently small to satisfy the adiabatic condition. However, for $\omega_1 = 0.1$, the system leaves the adiabatic limit. This results in a low fidelity over time, and the pumped energy loses its stable linear trend, breaking into erratic segments with fluctuating slopes, similar to the diabatic behavior for a single quantum spin shown in Martin et al. \cite{martin_topological_2017}.

\subsection*{Alternative setups}

\noindent The frequency conversion mechanism is not bound to a unilateral driving scheme where both incommensurate fields act simultaneously on the $N^{th}$ site. To demonstrate the flexibility of the approach, we investigated an alternative bilateral setup where the two phase drives act at opposite ends of the $XXZ$ chain. In this distributed protocol, the static and rotating fields are decoupled. The left boundary is subjected to a superposition of the static longitudinal field and a field rotating in the $xz$ plane. Conversely, the right boundary is driven solely by a field rotating in the $xy$ plane. The boundary terms in the Hamiltonian are thus modified to $\mathbf{B}_1(\theta_{1t}) = (1,0,0) + B(\cos{\theta_{1t}}, 0, \sin{\theta_{1t}})$ and $\mathbf{B}_N(\theta_{2t}) = B(\cos{\theta_{2t}}, \sin{\theta_{2t}}, 0)$. Just as in the single-sided configuration, the two independent driving phases span a torus, and the distributed driving protocol successfully generates a quantized topological energy pump. The ground state Chern number is again determined by the number of spectral degeneracy points enclosed by the torus. This alternative configuration highlights the nonlocal nature of the topological pumping, which can be used to transfer energy along the chain from one edge to the other.

\begin{figure}[t]
    \centering
    \includegraphics[width=0.99\linewidth]{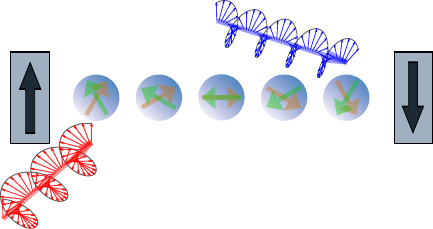}
    \caption{A general frequency converter setup with distributed boundary driving and fine tuning to the degeneracy point. A wide variety of driving protocols can be engineered by applying different combinations of static tuning fields to arbitrary sites to establish the degeneracy landscape, while subjecting other localized spins to the incommensurate phase drives.}
    \label{fig:Other_Setup}
\end{figure}

Furthermore, we can fine-tune the system close to a spectral degeneracy point using static boundary fields. E.g., we consider a configuration where the left and right ends of the chain are pinned by strong, anti-aligned static magnetic fields, $\mathbf{B}_1 = (1,0,0)$ and $\mathbf{B}_N = (-1,0,0)$ for odd lengths of chain. We then apply the two phase drives to any site $m$ ($1 \leq m \leq N$), governed by the local term $\mathbf{B}_{m}(\boldsymbol{\theta}_t) = B(\cos{\theta_{1t}}, \sin{\theta_{1t}}, 0) + B(\cos{\theta_{2t}}, 0, \sin{\theta_{2t}})$, see Fig.~\ref{fig:Other_Setup}. Because the global system is already tuned exactly to the quantum critical point by the boundary fields, even perturbatively small driving amplitudes $B$ enclose the singular point in the toroidal parameter space, yielding a Chern number of one, and induced topological power transfer.

More generally, the realization of the quantized topological pump depends on enclosed ground-state degeneracy points in the parameter space of quasi-periodic driving trajectories. As long as this topological criterion is satisfied, a wide variety of driving protocols can be engineered. This includes applying different combinations of static tuning fields to arbitrary sites to establish the degeneracy landscape, while subjecting other localized spins to the incommensurate phase drives. This structural flexibility broadens the experimental scope for designing topological frequency converters in the spin chains. 

\section{Conclusion and Outlook}\label{sec:Conclusion}

\noindent We establish a mechanism for generating arbitrarily large and tunable topological pumping in a quasiadiabatically driven interacting $XXZ$ spin chain applied to topological frequency conversion. The topological invariant is governed by the number of degeneracy points enclosed by the driving trajectory, where each degeneracy has topological charge one. These degeneracies are organized into universal, hyperbolic curves in the anisotropy-magnetic field phase space. Crucially, these curves accumulate systematically as the chain length increases, leading to a large Chern number for longer chains. This scaling behavior is primarily due to the interaction effects in the quantum many-body system, distinct from standard non-interacting topological pumps, where the Chern number is typically fixed by the band structure and restricted to low integer values. Furthermore, we demonstrate that this quantized conversion is robustly maintained even when the driving fields are spatially distributed across opposite ends of the chain, confirming that the frequency conversion is a global property of the entire chain rather than a localized edge effect. 

A possible solid-state realization of the setup involves atom-by-atom engineering using scanning tunneling microscopy (STM). In this context, chains of magnetic atoms (e.g., Fe or Ti) could be assembled on insulating substrates (such as MgO) to form Heisenberg spin chains with engineered exchange couplings \cite{khajetoorians_current-driven_2013, ternes_spin_2015,ChoiHeinrich_Atomic_2019,yang_probing_2021}. For our specific protocol, the boundary fields might be implemented using magnetic islands or tips \cite{schlenhoff_atomic_2012, khajetoorians_realizing_2011, khajetoorians_current-driven_2013}. Ideally, a static magnetic island placed near the first spin would exert a local Zeeman field via exchange interaction, pinning the first boundary spin ($B_1$). The dynamic driving at the other end could conceivably be achieved by applying time-varying radio-frequency (RF) fields through the STM tip or by modulating the tunnel current to induce spin-torque effects locally at the $N^{th}$ site \cite{baumann_electron_2015}.

Ultracold atoms in optical lattices might also simulate this effect, offering precise control over the exchange anisotropy $\Delta$. In this platform, where the spin states are encoded in internal hyperfine levels \cite{citro_thouless_2023} like the ones of ${}^7$Li \cite{jepsen_spin_2020}, the exchange anisotropy $\Delta$ could be tuned via Feshbach resonances or by exploiting the superexchange mechanism with polarization-dependent optical potentials \cite{trotzky_controlling_2010,chen_controlling_2011}. Because our setup utilizes the exchange anisotropy to control the Chern number, this tunable environment directly provides a mechanism to control the conversion ratio. Furthermore, the required localized boundary fields could be generated using focused laser beams (optical tweezers) or magnetic field gradients that address the chain edges individually \cite{weitenberg_single-spin_2011}. In such a setup, the quasi-adiabatic driving protocol would correspond to modulating the phases or amplitudes of these local addressing beams, effectively rotating the synthetic magnetic field vector seen by the boundary spin.
 
While the scalability of the Chern number with the system size offers a route to high-gain conversion, a challenging aspect could be respecting the adiabatic time scale, which increases asymptotically linearly with the system size in the thermodynamic limit \cite{alcaraz_conformal_1988}. Let us note that the linear scaling close to the ground state is actually beneficial compared to the exponential gap closure in the middle of the spectrum. Further stabilization could yet be necessary by employing optimized control techniques, such as shortcuts to adiabaticity \cite{Campo_ShortcutToAdiabacity_2013, Aslani_SuperpositionStatesViaShortcutToAdiabcity_2023}.

Moreover, the energy transfer observed in the distributed driving configuration implies a fundamental nonlocal transport mechanism across the bulk of the chain. Since power is injected at one boundary and extracted at the other, this transfer must be mediated by a physical current propagating through the chain. Characterizing this potentially quantized transport quantity, such as a topological energy or spin current, presents a compelling avenue for future investigation.

\begin{acknowledgments} 

\noindent We thank Pei-Xin Shen for helpful discussions. A.T.~and T.P.~acknowledge funding by the European Union (ERC, QUANTWIST, Project number 101039098). The views and opinions expressed are however those of the authors only and do not necessarily reflect those of the European Union or the European Research Council, Executive Agency. A.T.~and T.P.~acknowledge funding by the Cluster of Excellence ``CUI: Advanced Imaging of Matter'' of the Deutsche Forschungsgemeinschaft (DFG) – EXC 2056 – Project ID 390715994. M.T.~acknowledges support from the National Science Center (Poland) OPUS Grant No. 2021/42/B/ST3/04475, and the Foundation for Polish Science project ``MagTop'' (No. FENG.02.01-IP.05-0028/23) cofinanced by the European Union from the funds of Priority 2 of the European Funds for a Smart Economy Program 2021-2027 (FENG), and by the NAWA Bekker Grant No. BPN/BEK/2024/1/00310 (Poland).  
\end{acknowledgments}

\onecolumngrid 
\appendix
\section*{Appendix}
\noindent This appendix provides the theoretical framework and numerical data supporting the main text. First, we derive the topological power transfer relation for two incommensurate drives using first-order adiabatic perturbation theory within the Thouless pumping framework. Second, we present the comprehensive fitting parameters used to mathematically characterize the universal hyperbolic degeneracy manifolds across various spin chain lengths.

\section{Power transfer through quasi-adiabatic drives} \label{app:powerTransfer}
\noindent In this Appendix, we derive the power transferred by two incommensurate adiabatic drives acting on a Hamiltonian via the Thouless pumping mechanism. This formulation is equivalent to the transport formalism in synthetic Floquet lattices \cite{martin_topological_2017, crowley_topological_2019}. The time evolution of the Hamiltonian driven by the two periodic parameters $\mathbf{\theta} = (\theta_{1t}, \theta_{2t}) = (\omega_1t+\phi_1, \omega_2t +\phi_2)$ is 
\begin{align}
    H(\boldsymbol{\theta}_t) = & H(\theta_{1t}, \theta_{2t}), \text{\hspace{7em}   with}\\
    \frac{\partial H}{ \partial t} = & \frac{\partial H}{\partial \theta_1} \cdot \frac{\partial \theta_1}{\partial t} + \frac{\partial H}{\partial \theta_2} \cdot \frac{\partial \theta_2}{\partial t} \hspace{1em} =  \hspace{1em} \frac{\partial H}{\partial \theta_1} \cdot \omega_1 + \frac{\partial H}{\partial \theta_2} \cdot \omega_2.
\end{align}
We aim to calculate the work performed by the first drive, defined by the instantaneous power $\omega_1 \langle \partial_{\theta1} H\rangle$. In the adiabatic limit, considering $\hbar =1$, utilizing first-order perturbation theory and excluding the instantaneous phase, the time-evolved ground state is approximately \cite{thouless_quantization_1983}  
\begin{align}
    |\psi_0(t)\rangle&\approx|\psi_0^{in}(t)\rangle+i \sum_{n\neq0}\frac{\langle\psi_n^{in}(t)|\dot{\psi}_0^{in}(t)\rangle}{\epsilon_n(t)-\epsilon_0(t)}|\psi_n^{in}(t)\rangle,
\end{align}
where the superscript ``in" denotes the instantaneous eigenstates of $H(t)$, $\epsilon_n(t)$ is the energy of the $n$-th eigenstate, and $n=0$ denotes the gapped ground state. The instantaneous power transfer is expressed as   
\begin{align}
    \omega_1\langle\psi_0(t)|\frac{\partial H}{\partial \theta_1}|\psi_0(t)\rangle
    &=\omega_1 \left[\langle\psi_0^{in}(t)|\frac{\partial H}{\partial \theta_1}|\psi_0^{in}(t)\rangle
    +i \sum_{n\neq0}\frac{\langle\psi_0^{in}(t)|\frac{\partial H}{\partial \theta_1}|\psi_n^{in}(t)\rangle\langle\psi_n^{in}(t)|\dot{\psi}_0^{in}(t)\rangle}{\epsilon_n(t)-\epsilon_0(t)} + {\rm h.c.} \right] \\
    & = \omega_1 \left[ \langle\psi_0^{in}(t)|\frac{\partial H}{\partial \theta_1}|\psi_0^{in}(t)\rangle +i \sum_{n\neq0} \left\langle \frac{\partial\psi_0^{in}(t)}{\partial \theta_1}\Bigg|\psi_n^{in}(t) \right\rangle\langle\psi_n^{in}(t)|\dot{\psi}_0^{in}(t)\rangle + {\rm h.c.} \right] . \\
\end{align}
Substituting the relevant basis identities leads to \cite{tripathi_generalized_2025} 
\begin{align}
    \omega_1 \left\langle \frac{\partial H}{\partial \theta_1} \right\rangle &=  \omega_1 \frac{\partial \epsilon_0}{\partial \theta_1}  + i \left\langle \frac{\partial\psi_0^{in}(t)}{\partial \theta_1}\Bigg| \frac{\partial \psi_0^{in}(t)}{\partial \theta_2} \right\rangle \omega_1 \omega_2  + \underbrace{i \left\langle \frac{\partial\psi_0^{in}(t)}{\partial \theta_1}\Bigg|\frac{\partial \psi_0^{in}(t)}{\partial \theta_1} \right\rangle \omega_1^2 + {\rm h.c.}}_{=0}
\end{align}
The time-averaged power pumped is 
\begin{align}
    P = &\lim_{T \to \infty} \frac{1}{T} \int_{0}^{T} \left\langle \frac{\partial H}{\partial \theta_1} \right\rangle \omega_1 {\rm d}t \nonumber \\
    P = & \lim_{T \to \infty} \frac{1}{T} \int_{0}^{T} i \left(\left\langle \frac{\partial\psi_0^{in}(t)}{\partial \theta_1}\Bigg|\frac{\partial \psi_0^{in}(t)}{\partial \theta_2} \right\rangle - \left\langle \frac{\partial\psi_0^{in}(t)}{\partial \theta_2}\Bigg|\frac{\partial \psi_0^{in}(t)}{\partial \theta_1} \right\rangle\right) \omega_1 \omega_2 {\rm d}t.
\end{align}
Invoking the ergodic theorem for incommensurate frequencies $\omega_1$ and $\omega_2$ \cite{birkhoff_proof_1931}, the time integral over the quasi-periodic trajectory is equivalent to a surface integral over the entire toroidal parameter space,
\begin{align}
    P = &\frac{\omega_1\omega_2}{(2\pi)^2}\iint_{0}^{2\pi} i \left(\left\langle \frac{\partial\psi_0^{in}(t)}{\partial \theta_1}\Bigg|\frac{\partial \psi_0^{in}(t)}{\partial \theta_2} \right\rangle - \left\langle \frac{\partial\psi_0^{in}(t)}{\partial \theta_2}\Bigg|\frac{\partial \psi_0^{in}(t)}{\partial \theta_1} \right\rangle\right) {\rm d}\theta_1 {\rm d}\theta_2, \\
    P = & C \frac{\omega_1\omega_2}{2\pi}.
\end{align}
where $C$ is the Chern number defined in Eq.~(\ref{Eq:ChernNumber}). This result represents the power gained from the first drive. By energy conservation, the second drive loses power by the equivalent amount $-C\omega_1\omega_2/2\pi$, thereby realizing quantized topological frequency conversion. 

\section{Parameters of universal curves of quantum critical points} \label{app:FitParameters}
\begin{figure}[h]
        \centering
        \setlength{\unitlength}{0.1\textwidth}
        \begin{picture}(10,4)
            \put(0.5,0){\includegraphics[width=0.9\textwidth]{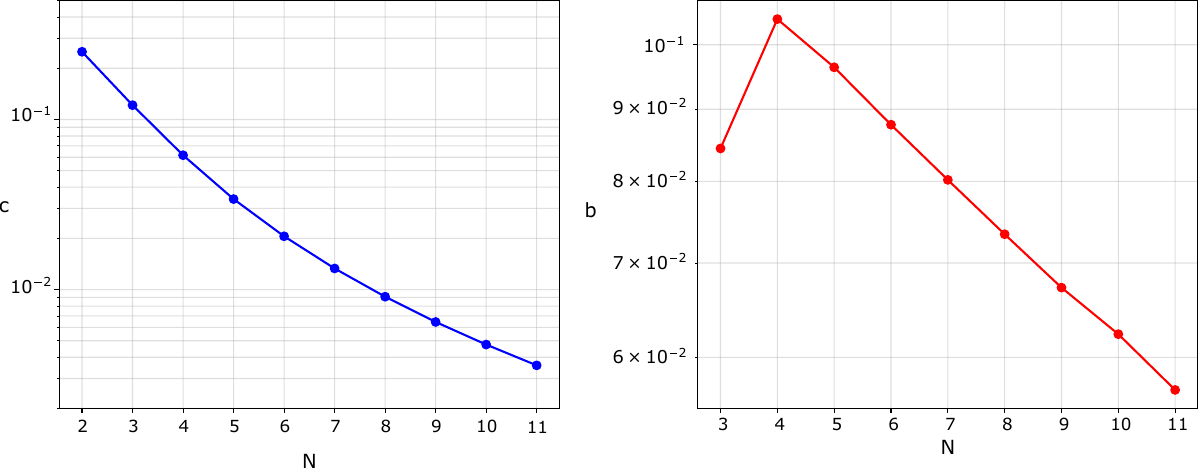}}
            \put(0.8,3.7){\textbf{(a)}}
            \put(5.6,3.7){\textbf{(b)}}
        \end{picture}        
    \caption{Scaling of the hyperbolic fit parameters with chain length N. (a) The parameter $c$ of the new accumulated trajectories with increase of $N$, plotted on a logarithmic scale. (b) The parameter $b$ (for $N \geq 3$ as $b=0$ for $N=2$), plotted on a logarithmic scale.}
    \label{fig:FitParameters}
\end{figure}

\noindent In this appendix, we provide the explicit fit parameters, $b$ and $c$, derived from the hyperbolic fits $(\Delta/J-\cos(\pi/N))(B+b) = c$ of the universal quantum critical points previously shown in Fig.~\ref{fig:PhaseSpace}. As the chain length $N$ increases, the critical trajectories accumulate. The extracted parameters for these emergent curves are summarized as a function of system size in Tab.~\ref{tab:Parameters}, and their scaling behaviors are illustrated in Fig.~\ref{fig:FitParameters}.

\begin{table}[h]
    \centering
    \begin{tabular}{|c||c|c|c|c|c|c|c|c|c|c|}
    \hline
    \textbf{N} & 2 & 3 & 4 & 5 & 6 & 7 & 8  & 9 & 10 & 11 \\ \hline
    \textbf{b} & 0.000000 & 0.084414 & 0.104276 & 0.096380 & 0.087751 & 0.080188 & 0.073374 & 0.067248 & 0.062321 & 0.056895 \\ \hline 
    \textbf{c} & 0.250000 & 0.121336 & 0.061636 & 0.034043 & 0.020572 & 0.013309 & 0.009081 & 0.006451 & 0.004750 & 0.003586 \\ \hline
    \end{tabular}
    \caption{Hyperbolic fit parameters $b$ and $c$ extracted from the quantum critical points for various spin chain lengths $N$ of Fig.~~\ref{fig:PhaseSpace}. The fits correspond to the relation $(\Delta/J-\cos(\pi/N))(B+b) = c$.}
    \label{tab:Parameters}
\end{table}
\end{document}